\documentclass{article}
\usepackage[utf8]{inputenc}
\usepackage{url}
\usepackage{graphicx,graphics}
\usepackage{appendix}

\title{A Computable Piece of Uncomputable Art\\whose Expansion May Explain the Universe in Software Space}
\author{Hector Zenil$^{1,2,3,4}$\\
$^1$Oxford Immune Algorithmics, Reading, U.K.;\\
$^2$The Alan Turing Institute, British Library, London, U.K.;\\
$^3$Algorithmic Dynamics Lab, Unit of\\ Computational Medicine, Center for Molecular\\Medicine, Karolinska Institute, Stockholm, Sweden; \\
$^4$Algorithmic Nature Group, LABORES for the Natural\\and Digital Sciences, Paris, France.\\
hector.zenil@cs.ox.ac.uk}
\date{}

\begin{document}

\maketitle

\begin{abstract}
At the intersection of what I call uncomputable art and computational epistemology, a form of experimental philosophy, we find a most exciting and promising areas of science related to causation with an alternative, possibly best possible, solution to the challenge of the inverse problem. That is, the problem of finding the possible causes, mechanistic origins, first principles, and generative models of a piece of data from a physical phenomenon. Here we explain how generating and exploring software space following the framework of Algorithmic Information Dynamics, it is possible to find small models and learn to navigate a sci-fi-looking space that can advance the field of scientific discovery with complementary tools to offer an opportunity to advance science itself.
\end{abstract}

\section{Failure is at the Origin of Computer Science}

At the beginning of the twentieth century and through the end of the Second World War, computers were human, not electronic. The work of a computer consisted in solving tedious arithmetical operations with paper and pencil. This was looked upon as work of an inferior order.

At an international mathematics conference in 1928, David Hilbert and Wilhelm Ackermann suggested the possibility that a mechanical process could be devised that would be capable of proving all mathematical assertions without need of human intervention, given the tedious and almost mundane nature of such a step-by-step undertaking without much human creativity involved.

This proposition is referred to as \textit{Entscheidungsproblem} in German, or `the decision problem' in English. If a human computer did no more than execute a mechanical process, it was not difficult to imagine that arithmetic would be amenable to a similar sort of mechanisation. The origin of Entscheidungsproblem dates back to Gottfried Leibniz, who having (around 1672) succeeded in building a machine based on the ideas of Blaise Pascal that was capable of performing arithmetical operations (named Staffelwalze or the Step Reckoner), imagined a machine of the same kind that would be capable of manipulating symbols to determine the truth value of mathematical principles. To this end
Leibniz devoted himself to conceiving a formal universal language, which he designated characteristica universalis, a language that would encompass, among other things, binary language and the definition of binary arithmetic.

In 1931, Kurt G\"odel~\cite{godel} concluded that Hilbert's intention (also referred to as `Hilbert's programme') of proving all theorems by mechanising mathematics was impossibe to realise -except in the case of trivial, constrained versions of mathematical frameworks  since the body of
mathematics needed to contain basic axioms such as those in arithmetic or set theory. G\"odel advanced formulas that codified arithmetical truths in arithmetical terms that could not be proved without arriving at a contradiction. Even worse, his formulas 
implied that there was no set of axioms that contained arithmetic free of unprovable true formulae, 
and that any extension to axioms 
would only lead to the same inconclusive statements in the theory.

In 1944, Emil Post, another key figure in the development of the concepts of
computation and computability (focusing especially on the limits of computation) found~\cite{post} that this problem was intimately related to one of the twenty-three problems (the tenth) that Hilbert, speaking at the Sorbonne in Paris, had declared the most important challenges for twentieth century mathematics.

Usually, Hilbert's programme is considered a failure, but not only it saved human creativity from full mechanisation but led to computation as a science by own virtue, perhaps even more than others that can be translated to finite or purely mechanistic procedures. Even though G\"odel debunked~\cite{godel} the notion that what was true could be proved, presenting a negative solution to the `decision problem' (Hilbert's tenth problem), followed by negative results based on G\"odel's~\cite{davis,robinson,matiyasevich}.

\subsection{All for one machine, and one machine for all}

Not long after G\"odel, Alan M. Turing made his appearance. Turing contemplated the problem of decision in much cruder terms. If the act of performing arithmetical operations is mechanical, why not substitute a mechanical device for the human computer? Turing's work represented the first abstract description of the digital general-purpose computer as we know it today. Turing defined what in his article he termed an `a' computer (for `automatic'), now known as a Turing machine.

A Turing machine is an abstract device which reads or writes symbols on a tape one at a time, and can change its operation according to what it reads, and move forwards or backwards through the tape. The machine stops when it reaches a certain configuration (a combination of what it reads and its internal state). It is said that a Turing machine produces an output if the Turing machine halts, while the locations on the tape the machine has visited represent the output produced.

The most remarkable idea advanced by Turing is his demonstration that there is an `a' machine that is able to read other `a' machines and behave as they would for an input s. In other words, Turing proved that it was not necessary to build a new machine for each different task; a single machine that could be reprogrammed sufficed. This erases the distinction between program and data, as well as between software and hardware, as one can always codify data as a program to be executed by another Turing machine and vice versa, just as one can always build a \textit{universal machine} to execute any program and vice versa.

Turing also proved that there are Turing machines that never halt, and if a Turing machine is to be \textit{universal}, and hence able to simulate any other Turing machine or computer program, it is actually imperative that it never halt
for an infinite number of inputs of a certain type (while halting for an infinite number of inputs of some other type). Indeed this is what Turing would have been led to expect, given G\"odel's results and what he wanted to demonstrate: that Hilbert's mechanisation of mathematics was impossible. This result is known as the undecidability of the halting problem.

In his seminal article Turing defined not only the basis of what we today know as digital general-purpose computers, but also software, programming and subroutines. And thus without a doubt it represents the best answer to date that we have to the question `What is an algorithm?'

In fact in Alan Turing's work on his universal machine, he even introduced the concept of a subroutine that helped him in his machine construction. These notions are today the cornerstone of the field that Turing, more than anyone else, helped establish, viz. Computer Science.

\section{A Slice of the Uncomputable Universe}

Once we approach the problem of defining what an algorithm is and arrive at the
concept of \textit{universality} that Turing advanced, the question to be considered in greater detail concerns the nature of algorithms. Given that one now has a working definition of the algorithm, one can begin to think about classifying problems, algorithms and computer programs by, for example, the time they take or the storage memory they may require to be executed. One may assume that the time required for an algorithm to run would depend on the type of machine, given that running a computer program on a Pentium PC is very different from executing it on a state-of-the-art super computer. This is why the concept of the Turing machine
is so important---because any answers to questions about problems and algorithmic resources will only make sense if the computing device is always the same. And that device is none other than the universal Turing machine. So for example, every step that a Turing machine performs while reading its tape is counted as a time step.

Many algorithms can arrive at the same conclusion by different paths, but some may be faster than others. This is now a matter that's  carefully considered  when fixing the framework for Turing's model of computation: one asks whether there is an algorithm that surpasses all others in economy as regards the resources required, when using exactly the same computing
device. These are the questions that opened up an entire new field in the wake of Turing's work, the development of which Turing would certainly have been delighted to witness. This field is today referred to as the theory of Computational Complexity, which would not have been possible without a concept such as that of the universal Turing machine. The theory of Computational Complexity focuses on classifying problems and algorithms according to the time they take to compute when larger inputs are considered, and on how size of input and execution time are related to each other. This is all connected to two basic resources needed in
any computation: space (storage) and time. For example, one obvious observation relating to this theory is that no algorithm will need more space than time to perform a computation. One
can then quickly proceed to ask more difficult but also more interesting questions, such as whether a machine can execute a program faster if it is allowed to behave probabilistically instead of  deterministically. What happens if one adds more than one tape to a universal Turing machine operation? Would that amount to implementing an algorithm to solve a problem much faster? Or one may even ask whether there is always an efficient alternative to every inefficient algorithm, a question that may lead us to a fascinating topic connecting computers and physics.

\begin{figure}[ht]
\centering
\includegraphics[width=11cm]{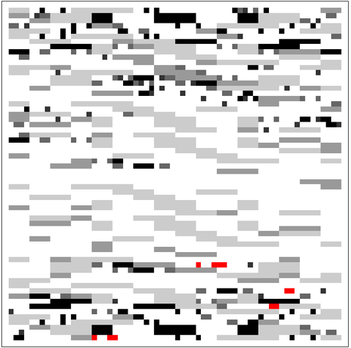}
\caption{\label{uncomputableart1} Visualising the Computational Universe: Runtime Space in a Peano Curve''~\cite{zenilruntimes} by this author. 1st. prize in the Computational Imagery category of the ``2011 Kroto Institute Scientific Image Competition'' held in the UK.}
\end{figure}

\begin{figure}[ht]
\centering
\includegraphics[width=11cm]{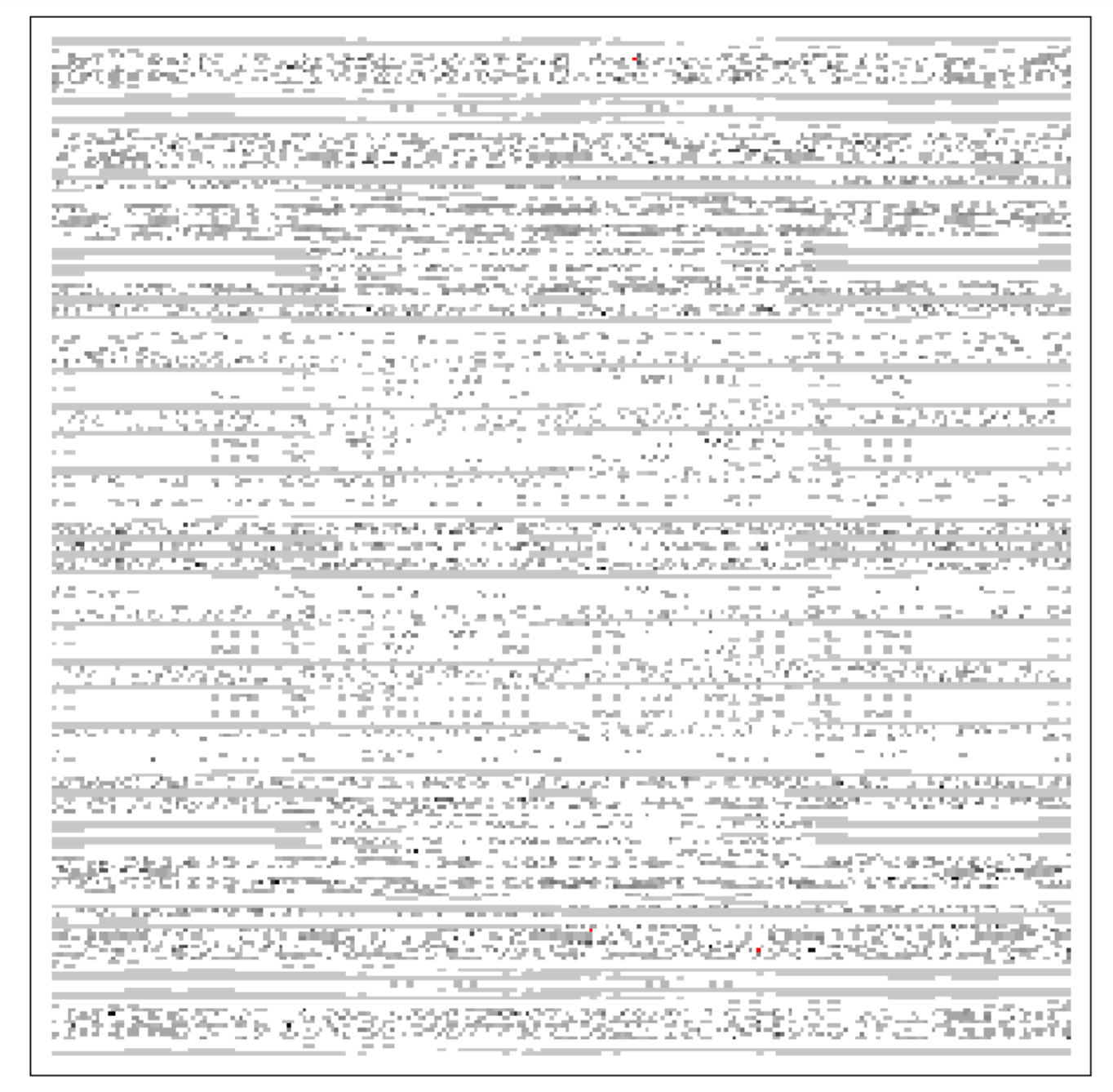}
\caption{\label{uncomputableart2} Runtime deep field of a segment of runtimes from 7\,529\,536 Turing machines
in (3,2). The (3, 2) Busy Beavers are barely visible as isolated red points (online and colour printed versions only).}
\end{figure}

A full technical explanation of figure ~\ref{uncomputableart2} is provided in~\cite{zenilruntimes}. My results were later used to validate and produce some other results, such as XXX these papers (that have cited mine! including Calude's!).

Each pixel represents a computer program with empty input running on an abstract computer (a Turing machine). The colour indicates the runtime of the program upon completion of its computation. The darker a pixel the longer the program took to produce its output. White pixels are computer programs that do not produce any output because they never halt (just as happens when your computer stalls forever, sometimes prompting you to reboot).

Red pixels are the computer programs that take the longest time relative to other programs of the same size before finishing (they are also called Busy Beavers). I call this colour type coding the computability spectrum. Programs were sorted by size from smallest to largest (in terms of number of instructions) in such a way that one could be certain that any program would eventually be reached (including versions of MS Windows and Angry Birds that can be encoded as one-run computer programs).

Because there is no upper limit to the size of programs, the space of all computer programs is infinite but discrete, perhaps mirroring our own world. Because an enumeration of computer programs is linear (you start with program 1, then proceed to program 2, and so on), pixels depicted in this 2-dimensional figure are distributed along a ``space-filling'' Peano curve. This is a curve that guarantees that 2 contiguous elements in the enumeration remain as close as possible to each other in the 2D configuration. 

The picture has the striking particularity that it is ultimately uncomputable. Imagine you point a telescope and take a picture of an object in the computational universe for which some pixel values cannot be determined, not only in practice but in principle-- that is, no matter how good your camera, or how many 'terapixels' some future version of it may have. This slide of the computational universe was only possible because it is a tiny region of small computer programs for which all pixels can be determined by zooming in deep enough.

However, G\"odel and Turing proved that if you zoomed out to a larger view of the computational universe, the picture would begin to look completely blurry because pixel colours would not be determinable. In other words this picture is the inverse equivalent of Hubble's Deep Field space--the larger the space field, the blurrier.

Just as our access to the visible universe is determined by the speed of light, due to similar limitations we can only see a small part of the computational universe, limitations both physical and computational (it seems we cannot build more powerful computers than the ones we already have to overcome these limitations). Hence, most of the computational space will remain unknowable, not only due to a limitation of resources but due to a fundamental barrier originating
in the digital world and extending into our own reality.

Only incremental improvements to this picture are possible. While the picture is ultimately uncomputable and unattainable, one can make an informed guess as to what it would look like if we zoomed out. Theory predicts the average colour of the picture at a greater scale, the theory in question being the theory of algorithmic probability, closely related to what is known as the Chaitin $\Omega$ number, or the halting probability. It turns out that the picture will be mostly white, as in an ever-expanding universe finding computer programs that halt and produce something will be increasingly harder. Indeed, most programs will never halt, and those that do will do so relatively quickly in proportion to their size.

\section{How to Navigate in Software Space}

In his seminal book~\cite{wolfram}, Wolfram introduced the concept of the software space int he context of mining it to find explanations of the physical world. The software space is, however, infinite, unreachable and full of landmines, because of uncomputability and irreducibility~\cite{empirical}, and any attempt to traverse it will hugely depend on the enumeration chosen leading to biases in its exploration some of which, for example, can be seen in figures~\ref{uncomputableart1}, and ~\ref{uncomputableart2} where some regions share more levels of grey compared to others.  

First, it is important to recognise that a positive result of Turing universality is that there is only, and one only, software space, but many ways to generate it and navigate it.  Indeed, to generate software space is to navigate it and the other way around and it all depends on the computing model and the enumeration of the computer programs that populate the universe, that is the order in which they are generated and navigated.  Because of Turing universality, we know that no matter how we generate or navigate the space, the same programs will be found in any simulated space, and each space is just a different representation or simulation of another one and the same.  However, this does not mean that one can generate and explore programs of interest at the same rate.  In some models, this can be slower than others by a simple enumeration reshuffling.  Functional programming languages, for example, will tend to generate recursive functions, the type of functions that have some preference in mathematics and produce regular algorithmic or statistical patterns, more often and quicker than say, an enumeration based on rule-states (like Turing machines) where, for example, states that are not used, or rules from the product of machine symbols and states, will dominate certain local regions which if were not interesting will remain non-interesting and the other way around.

So, how to navigate the space to find anything interesting in the sense of matching data and model to solve inverse problems?  And why would we wish to navigate the space in the first place?  Along the lines of Wolfram~\cite{wolfram}, software space is an opportunity because computer programs in software space have the opportunity to represent an explanation of computable phenomena.  This is not necessarily something completely new, in fact, all human scientific exploration so far has been based on the idea of finding computable explanations that we write down in the form of computable formulae or in computable simulations running on Turing-equivalent computers. What is new and introduced by Wolfram is its systematic exploration as an opportunity, and the discovery of its richness given that, under some strong but reasonable assumptions, such as the (physical version of the) Church-Turing thesis, one can expect to find many models~\cite{smalldata}, if not all, to any piece of empirical data. So, actually finding these small models offers an opportunity to look for discrete models of real-world phenomena to solve inverse problems. What had not been attempted and framed were strategies to navigate in such space to reduce the exploration, quantify, and reuse the findings in a clever way, given the (infinite) computational resources needed to computably approximate the uncomputable.

In~\cite{mutations}, for example, we have found, an exploration that explains some aspects of biological evolution, including modularisation (e.g. genes), production of high-level structure, and pathways of mutation (e.g.leading to cancer).  Other applications of ours have ranged from thermodynamics~\cite{thermo}, to finance~\cite{finance}, cognitive sciences~\cite{ploscompbio}, molecular biology~\cite{nar}, and agriculture~\cite{gaucherel}; but a long list of uses from independent groups has also been recorded, ranging from psychometrics to  chemistry, and cancer research.

\subsection{The beautiful theory of Algorithmic Probability}

Algorithmic probability tells us that most programs will produce an output following exponential decay ($- \log_2$ $Pr(s)$). Hence most of them will produce the same strings and a very few will produce algorithmically random digital sequences. In other words, among the computer programs that do produce an output, most will be highly ordered, as if they were creating a structured digital universe of highly ordered patterns out of nothing (just like ours). Some of these programs that produce the same output may do so faster than others, but precisely how much faster a program can be is another open foundational question, and one of the greatest challenges in mathematics and computer science. If it were possible to find a program considerably faster than any other, then everything could be computed in `reasonable' time, regardless of the apparent difficulty of the computation.

In a later follow-up piece~\cite{naturalcomputing}, a connection between Turing machines and Turing patterns is made, explaining how symmetry-breaking creates structure---as opposed to randomness---out of nothing, with only computability being assumed.

At the centre of my ideas for guiding searches of interesting regions and programs as models in the computational universe, in this deep field view of the software space, are some methods I have introduced based on or inspired by the seminal concept of Algorithmic Probability, itself introduced by Solomonoff~\cite{solomonoff} and Levin~\cite{levin} and having affinities with  Chaitin's $\Omega$ number or halting probability~\cite{chaitin}. This concept of algorithmic probability is formally defined as follows,

$$AP(s) = \sum_{p:U(p) = s} 1/2^{|p|} $$

That is, the sum over all the programs $p$ for which a universal Turing machine $U$ outputs $s$ and halts.

The notion behind $AP$ is very intuitive. If one wished to produce the digits of $\pi$ randomly, one would have to try time after time until one managed to hit upon the first numbers corresponding to an initial segment of the decimal expansion of $\pi$. The probability of success is extremely small: $1/10$ digits multiplied by the desired quantity of digits. For example, $1/10^{2400}$ for a segment of 2400 digits of $\pi$. But if instead of shooting out random numbers one were to
shoot out computer programs to be run on a digital computer, the result would be very different.

A program that produces the digits of $\pi$ would have a higher probability of being produced by a computer. Concise and known formulas for $\pi$ could be implemented as short computer programs that would generate any arbitrary number of digits of $\pi$.

\subsection{Simplicity is likeliness and likeliness simplicity}

The length of the shortest program that produces a string is today the mathematical definition of randomness, as introduced by Kolmogorov~\cite{kolmo}, Chaitin~\cite{chaitin}, and Solomonoff~\cite{solomonoff}, and later expanded by Levin~\cite{levin} and also called Kolmogorov-Chaitin complexity. 

The idea is relatively simple. If a string $s$ of length $|s|$ cannot be produced by a program $|p|$ shorter than $|s|$ in bits, then the string $s$ is random because it cannot be effectively described in a shorter way than by $s$ itself, there being no program $p$ that generates $s$ whose length is shorter than $s$. Formally,

$$C_U(s) = \min\{|p|:U(p)=s\}$$

\noindent where $U$ is a universal optimal Turing machine. The invariance theorem~\cite{kolmo,chaitin,solomonoff} guarantees that such an optimal universal Turing machine exists for which the value of $C$ nuder different Turing-complete language implementations over the same object is bounded by a `small' constant. Formally, if $U_1$ and $U_2$ are such implementations and $C_{U_1}(s)$ and $C_{U_2}(s)$ are the values of algorithmic complexity of $s$ for $U_1$ and $U_2$ respectively, there exists an optimal $U$ and constant $c$ such that

$$|C_{U_1}(s) - C_{U_2}(s)| < c$$

Clearly, the longer the string, the less relevant is $c$, and the more stable the algorithmic complexity values.
One of the disadvantages is that, given the halting problem for Turing machines, $C$ is not computable, which is to say that given a string, there is no algorithm that returns the length of the shortest computer program that produces it. Efforts have been taken before to approximate it through statistical compression algorithms~\cite{zenilreview} but these fail at characterising algorithmic complexity in full and are rather implementations of variations of Shannon entropy~\cite{zkpaper,zenilreview2}.

Algorithmic probability and algorithmic complexity $K$ are inversely proportional, as established by the so-called algorithmic Coding theorem~\cite{cover,calude}:

$$|-\log_2 AP(s) - C(s)| < c $$

\noindent where $c$ is a constant independent of $s$. The Coding theorem implies that the algorithmic complexity can be estimated from the frequency of a string. In other words, complexity is in inverse relation to algorithmic probability.

To illustrate the above, let us consider $\pi$. Under the assumption of Borel's absolute normality of $\pi$, whose digits appear randomly distributed, and with no knowledge of the deterministic source and nature of $\pi$ as produced by short mathematical formulae, we ask how an entropy versus an algorithmic metric performs. First, the Shannon entropy rate (thus assuming the uniform distribution along all integer sequences of $N$ digits) of the first $gN$ digits of $\pi$, in any base, would suggest maximum randomness at the limit. However, without access to or without making assumptions as regards the probability distribution, approximations to algorithmic probability would assign $\pi$ high probability, and thus the lowest complexity, by the Coding theorem, as has been done in~\cite{d4,d5,kolmo2d,bdmpaper}.

Just as with $\pi$ but in application to graphs and networks, it has been proven how certain graphs can be artificially constructed to target any level of Shannon entropy~\cite{zkpaper,morzy}, preserving low algorithmic randomness.

But how relevant is the algorithmic Coding theorem in explaining, e.g., natural phenomena, if it only applies to Turing-universal systems? We know that the natural world allows and can carry out Turing-universal computation because we have been able to build computers that take elements from nature and make them perform as general-purpose computers. However, we don't know how much of the natural world is Turing-computable or how physical laws may naturally implement any form of computation. 

So, in~\cite{liliana} we showed that up to 60\% of the bias found in the output of systems that are not Turing universal may be explained by the algorithmic Coding theorem. This means that this theorem is far more relevant than expected, as it not only explains the way data and patterns distribute for a very particular computational model, but does so, to a considerable extent, for basically any computing model. It is not difficult to imagine that nature operates at some of these levels, perhaps not even on a fixed one but at multiple scales, with the algorithmic Coding theorem being relevant to all of them.

\subsection{My own navigation through software space}

\begin{figure}[ht!]
\centering
\includegraphics[width=6cm]{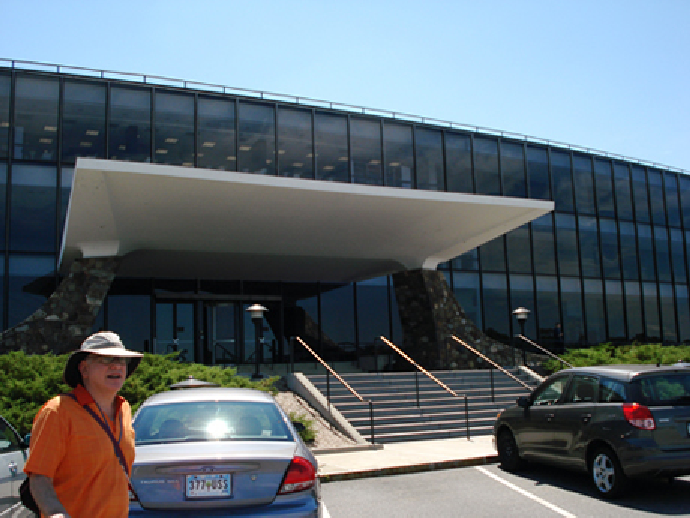}\\

\bigskip

\includegraphics[width=3.8cm]{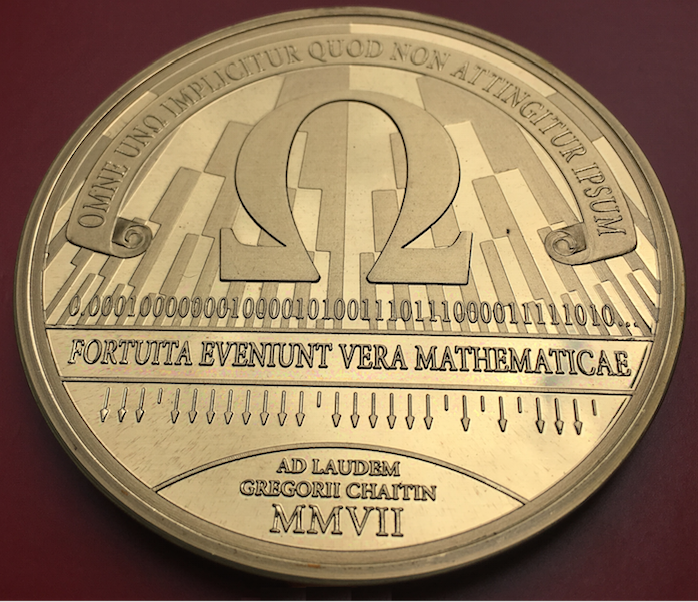}\hspace{.7cm}\includegraphics[width=4.05cm]{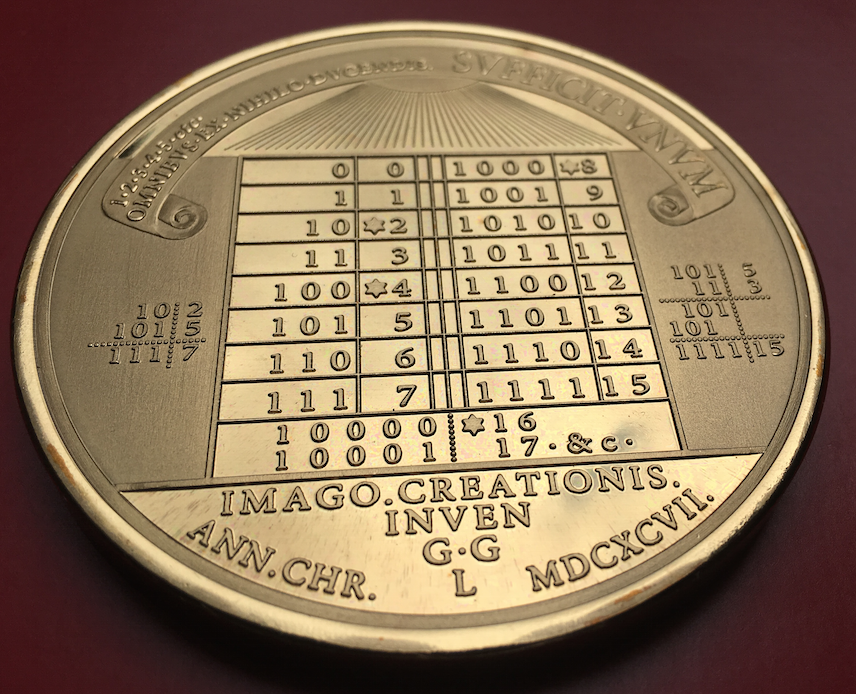}\\

\caption{(Top) A picture of Greg Chaitin I took outside his longtime office at IBM Research headquarters, the Thomas J. Watson Research Center. (Bottom) The two sides of the medal I helped Wolfram design, featuring material relating to Chaitin's life achievements on the obverse and Leibniz' own medal celebrating binary arithmetic on the reverse.}
\label{medal}
\end{figure}

The work of Greg Chaitin has been at the centre of my interests. Not only was he one of the examiners for my PhD in computer science in 2011, but my work would have not been possible without some of his theoretical work. I met Greg in 2006, having been invited to his home and to his office at IBM's legendary Thomas J. Watson Research Center (see figure ~\ref{medal}), both in Yorktown Heights, NY, U.S. His house was typical of an intellectual, full of objects and books piled one on top of the other. But Greg also had exotic pieces of sculpture scattered around, and an over-representation of Leibniz- and G\"odel-related books, the two major influences on his work and thinking.

\begin{figure}[h!]
\centering
\includegraphics[width=8cm]{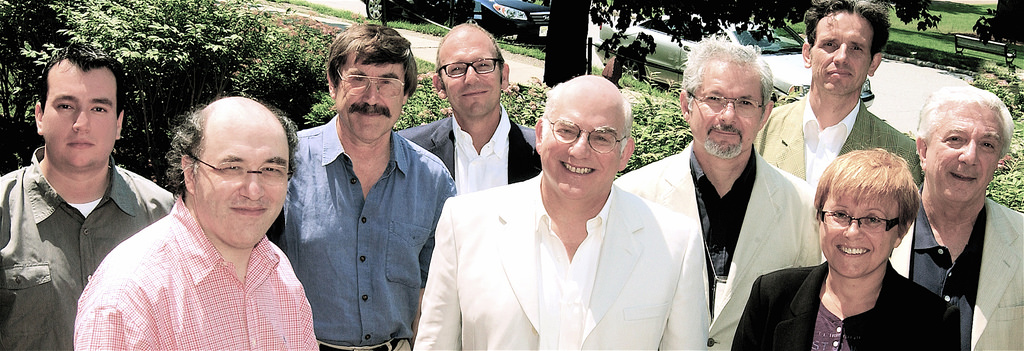}\\
Panel Part I: Is the universe random?\\

\bigskip

\includegraphics[width=8cm]{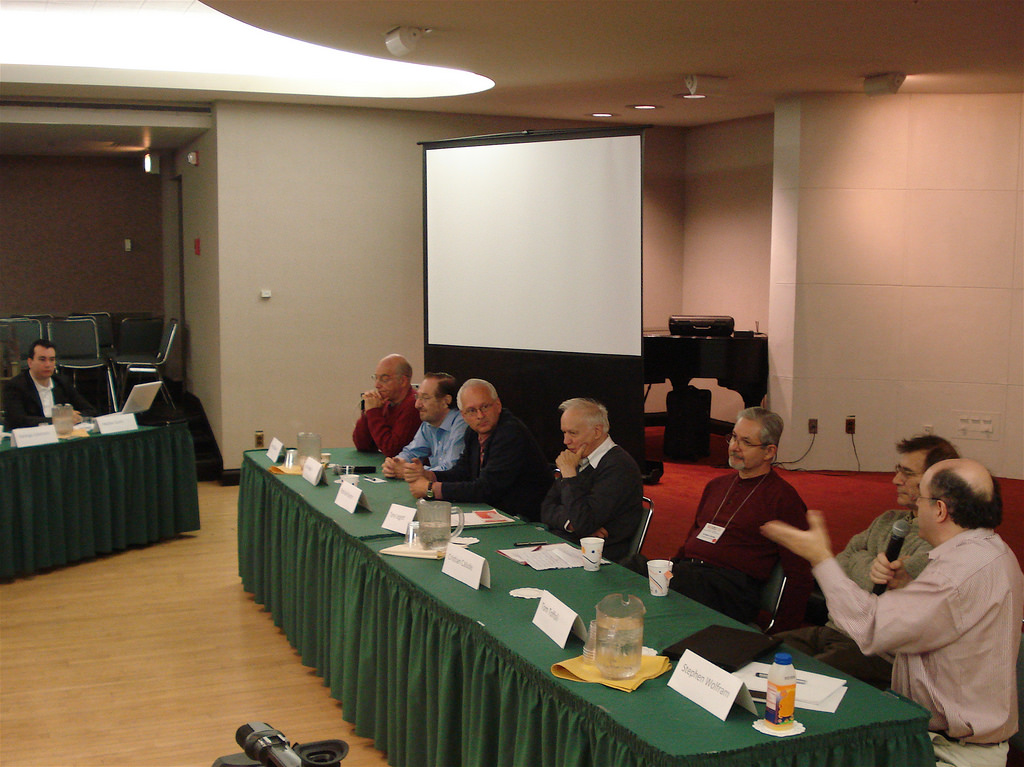}\\
Panel Part II: What is computation and How does nature compute?\\

\caption{Part I and Part II panel discussion pictures. (Top) From left to right: Hector Zenil, Stephen Wolfram, Paul Davies, Ugo Pagallo, Greg Chaitin, Cris Calude, Karl Svozil, Gordana Dodig Crnkovic, and John Casti. (Bottom) From left to right: Hector Zenil, Greg Chaitin, Ed Fredkin, Rob de Ruyter, Tony Legett, Cris Calude, Tom Toffoli, and Stephen Wolfram. Transcriptions in ~\cite{randomness} and ~\cite{computableuniverse}.}
\label{panels}
\end{figure}

In 2008, I organised a two-part panel discussion (figure~\ref{panels}) during the Wolfram Science conference at the University of Vermont in Burlington, an event that, together with a second meeting that I organised (with Adrian German) in 2008 at Indiana University Bloomington, will, I believe, come to be regarded as events of significant historical importance in the discussion of the ideas of the late 20th and early 21st centuries around questions of the analogue and the digital. The first part of the panel discussion addressed the question ``Is the Universe random?" Participating were Cris Calude, John Casti, Greg Chaitin, Paul Davies, Karl Svozil and Stephen Wolfram. The second part addressed the question ``What is computation and how does nature compute?". The participants were Ed Fredkin, Greg Chaitin, Tony Legett, Cris Calude, Rob de Ruyter, Tom Toffoli, and Stephen Wolfram, with George Johnson, Gerardo Ortiz and myself serving as moderators. Transcriptions of both parts were published in my volume~\cite{randomness} and ~\cite{computableuniverse}.

In 2007, I helped Stephen Wolfram design a commemorative medal to celebrate Gregory Chitin's 60th birthday, which also involved minting for the first time a 300 year old medal originally designed by Gottfried Leibniz to celebrate the invention or discovery of binary arithmetic (see figure ~\ref{medal}). I published a blog post soon after we came up with the idea for the medal, explaining the pre- and 
post-minting story of the medal~\cite{blogpost}.

At the centre of my research is Greg Chaitin's work on what is known as algorithmic probability, which in turn is related to  Chaitin's Omega ($\Omega$) number, also called the halting probability. My very first research paper~\cite{zenilcalude} was published in Chaitin's 60th birthday festchrift~\cite{calude}. This paper provided the first numerical evidence of a bias present in nature, a claim that I later advanced in a contribution that won an FQXi prize~\cite{zenilessay} in a competition on the subject of the digital or analogue nature of the universe. Such claims have more recently been picked up and expanded upon~\cite{kamal}--efforts in which my own methods have again played an important role--and have also been further explored by myself in the context of scientific discovery~\cite{maininfo,nar,nmi}.

\subsection{The Coding Theorem Method (CTM)}

The algorithmic \textit{Coding Theorem Method} (CTM)~\cite{d4,d5} provides the means for approximation via the frequency of a string. Now, why is this so? The underlying mathematics originates in the relation specified by algorithmic probability between frequency of production of a string from a random program and its algorithmic complexity. It is also therefore designated the algorithmic \textit{Coding theorem}, in contrast to another well known coding theorem in classical information theory. Essentially, the numerical approximation hinges on the fact that the more frequently a string (or object) occurs, the lower its algorithmic complexity. Conversely, strings with a lower frequency have higher algorithmic complexity. Otherwise stated,

$$CTM(s) = - \log_2 AP(s) + c$$

The way to implement a compression algorithm at the level of Turing machines, unlike popular compression algorithms which are heavily based on Shannon entropy, is to go through all possible compression schemes. This is equivalent to traversing all possible programs that are a compressed version of a piece of data, which is exactly what the CTM algorithm does.

\subsection{The Block Decomposition Method (BDM)}

Our approach to Chaitin's halting probability and Solomonoff-Levin's algorithmic probability consists in asking after the probability of a matrix being generated by a random Turing machine on a 2-dimensional array, also called a \textit{termite} or \textit{Langton's ant}~\cite{langton}. Hence an accounting procedure is performed using Turing machines that aims to approximate the algorithmic complexity of the structures identified. This technique is referred to as the \textit{Block Decomposition Method} (BDM), as introduced in~\cite{bdmpaper}. The BDM technique requires a partition of the adjacency matrix corresponding to the graph into smaller matrices. With these building blocks at hand we numerically calculate the corresponding algorithmic probability by running a large set of small 2-dimensional deterministic Turing machines, and then--by applying the algorithmic Coding theorem, as discussed above--its algorithmic complexity. 

Following such a divide-and-conquer approach we can then approximate the overall complexity of the original adjacency matrix by the sum of the complexity of its parts. Note that we have to take into account a logarithmic penalisation for repetition, given that $n$ repetitions of the same object only add $\log n$ to its overall complexity, as one can simply describe a repetition in terms of the multiplicity of the first occurrence. Technically, this translates into the algorithmic complexity of a labelled graph $G$ by means of $BDM$ which is defined as follows:

\begin{equation}
\label{newecaeq}
BDM(G,d) = \sum_{(r_u,n_u)\in A(G)_{d\times d}} \log_2(n_u)+K_m(r_u)
\end{equation}
where $K_m(r_u)$ is the approximation of the algorithmic complexity of the sub-arrays $r_u$ arrived at by using the algorithmic Coding theorem, while $A(G)_{d\times d}$ represents the set with elements $(r_u,n_u)$, obtained by decomposing the adjacency matrix of $G$ into non-overlapping squares, i.e. the block matrix, of size $d$ by $d$. In each $(r_u,n_u)$ pair, $r_u$ is one such square and $n_u$ its multiplicity (number of occurrences). From now on $BDM(G,d=4)$ will be denoted only by $BDM(G)$, but it should be taken as an approximation to $C(G)$ unless otherwise stated (e.g. when taking the theoretical true $C(G)$ value). Once CTM is calculated, BDM can be implemented as a look-up table, and hence runs efficiently in linear time for non-overlapping fixed size submatrices.

\section{Algorithmic Information Dynamics, a software calculus}

Unlike most complexity measures
~\cite{zenilreview} that are designed for static objects, except those related to dynamical systems (e.g. inverse problems with differential equations, Lyapunov exponents, etc), the measure of algorithmic complexity we introduced before can be adapted to characterise how dynamical systems move in software space by looking at the rate of change of algorithmic complexity of an object evolving over time. This measure is universal in the sense that it can deal with any computable feature that a system may display over time, either spontaneously or as a result of an external perturbation/intervention/interaction. Indeed, Algorithmic Information Dynamics is the combination of algorithmic probability and perturbation analysis in software space.

At the core of Algorithmic Information Dynamics~\cite{maininfo,crc}, the algorithmic causal calculus that we have introduced is the quantification of the change of complexity of a system under natural or induced perturbations, particularly the direction (sign) and magnitude of the difference of algorithmic information approximations denoted by $C$ between an object $G$, such as a cellular automaton or a graph,  and its mutated version $G^\prime$, e.g. the flip of a cell bit (or a set of bits) or the removal of an edge $e$ from $G$ (denoted by $G\backslash e= G^\prime$). The difference $| C(G) - C(G\backslash e) |$ is an estimation of the shared algorithmic mutual information of $G$ and $G\backslash e$. If $e$ does not contribute to the description of $G$, then $| C(G) - C(G\backslash e) | \leq \log_2|G|$, where $|G|$ is the uncompressed size of $G$, i.e. the difference will be very small and at most a function of the graph size, and thus $C(G)$ and $C(G\backslash e)$ have almost the same complexity. If, however, $| C(G) - C(G\backslash e) | \leq \log_2|G|$ bits, then $G$ and $G\backslash e$ share at least $n$ bits of algorithmic information in element $e$, and the removal of $e$ results in a loss of information. In contrast, if $C(G) - C(G\backslash e) > n$, then $e$ cannot be explained by $G$ alone, nor is it algorithmically not contained/derived from $G$, and it is therefore a fundamental part of the description of $G$, with $e$ as a generative causal mechanism in $G$, or else it is not part of $G$ but has to be explained independently, e.g. as noise. Whether it is noise or part of the generating mechanism of G depends on the relative magnitude of n with respect to $C(G)$ and to the original causal content of $G$ itself. If $G$ is random, then the effect of $e$ will be small in either case, but if $G$ is richly causal and has a very small generating program, then $e$ as noise will have a greater impact on $G$ than would removing $e$ from an already short description of $G$. However, if $| C(G) - C(G\backslash e) | \leq \log_2 |G|$, where $|G|$ is, e.g., the vertex count of a graph, or the runtime of a cellular automaton, $G$, then $e$ is contained in the algorithmic description of $G$ and can be recovered from $G$ itself (e.g. by running the program from a previous step until it produces $G$ with $e$ from $G\backslash e$).

We have shown, for example, how we can infer and reconstruct space-time evolutions via quantification of the disruption of a perturbation~\cite{maininfo} and even de novo~\cite{frontiers}, including initial conditions. From there, it can then be extracted a generating mechanism from the ordered time indices, from least to most disruptive, and produce candidate generating models. Simpler rules have simpler hypotheses, with an almost perfect correspondence in row order. Some systems may appear more disordered than others, but locally the relationship between single rows is for the most part preserved (indicating local reversibility). 

We have also shown that the later in time a perturbation is injected into a dynamical system, the less it contributes to the algorithmic information content of the overall space-time evolution~\cite{maininfo}. We then move from discrete 2D systems to the reconstruction of phase spaces and space-time evolutions of $N$-dimensional, continuous, chaotic, incomplete and even noisy dynamical systems.

Algorithmic Information Dynamics suggests that by performing perturbations to data and systems and quantifying the changes that these produces on the data models in software space, we can navigate software space to better understand a system or data in physical space.

\subsection{What the Universe may be made of}

Could it be that the physical universe is made or comes from a computational process and is therefore within software space? This would make our work even more relevant if computable model-driven explanations are approachable with the methods we have advanced allowing some degree of navigation in the boundless infinite software space.  Among the many interesting questions is how common computation universality is in software space, if it is very common one can expect both forward and inverse problems more difficult than expected. First, because the forward problem means that a universal model would be capable of arbitrary output making thus initial configurations and interactions with other systems more important.  For the inverse problem, that also means that matching model to data is more difficult because the relationship is between even larger sets than possibly imagined, as many more models will be able to match many more data, again making specific parameters all the more important.

What we have found~\cite{riedelzenil}, is that indeed, universality seems to be quite common. For this, we came up with an approach~\cite{dynamic} able to produce statistical evidence towards universality which is rather strong, to some extent even stronger than simple computation universality, a stronger version called intrinsic universality consisting on not only being able to emulate any computable function but also any computable function of the type and scale of the emulator itself.  And this is what we found, that most computer programs were capable of emulating both quantitatively and qualitatively most, or all, other computer programs in their own model space.  We did this by looking at the set of ever growing compilers (computable translations), that coupled with the original computer program would allow the program to compute quantitatively and qualitatively any other computer program in its own rule space and thus beyond its rule space assuming such evidence amounted to intrinsic universality. 

Our compilers (and compilers in general) are nothing but small finite computer programs that cannot be themselves universal because only operate once a computation has started and do not interact with the computation itself,

A positive result (if pervasive universality is read as a negative one, depending on the angle), is that we found also the a counter balance in the equation.  Not all compilers where easily found for all rules to be equally versatile, there was a strong inequality indicating that some programs are more naturally programmable than others, thus reintroducing an asymmetry compatible with other measures of complexity, namely resource (time) complexity.  Pervasive universality also means that software space is richer than possibly first imagined and may indicate that the real-world is equally rich (and thus, difficult to match) but also fascinating and with the possibility to explain the universe itself~\cite{comprehension,wpp}.

The methods and framework that we have introduced, Algorithmic Information Dynamics (AID)~\cite{scholarpedia}, provides tools to attempt to find those matches~\cite{nmi,maininfo,riedelzenil2}, even between sets of data and sets of models, together with a ranking rule to gauge such models based on algorithmic likeliness.  For this reason, we believe AID is a first step in the right direction.


\begin{thebibliography}{}

\bibitem{calude} C. Calude (ed), \textit{Randomness and Complexity: From Leibniz to Chaitin}, World Scientific Pub Co, Singapore, 2007.

\bibitem{davis} M. Davis, Hilbert's Tenth Problem is Unsolvable, \textit{American Mathematical Monthly}, 80: pp. 233--269, 1973.

\bibitem{robinson} M. Davis, Y.V. Matiyasevich, J. Robinson, Hilbert's Tenth Problem: Diophantine Equations: Positive Aspects of a Negative Solution. In Felix E. Browder (ed.). Mathematical Developments Arising from Hilbert Problems. \textit{Proceedings of Symposia in Pure Mathematics}, XXVIII.2. American Mathematical Society. pp. 323--378, 1976.

\bibitem{zenilcalude} J-P. Delahaye and H. Zenil, On the Kolmogorov-Chaitin complexity for short sequences. In C. Calude (ed) \textit{Randomness and Complexity: From Leibniz to Chaitin}, World Scientific Publishing Company, 2007.

\bibitem{d4} J.-P. Delahaye and H. Zenil, Numerical Evaluation of the Complexity of Short Strings: A Glance Into the Innermost Structure of Algorithmic Randomness, \emph{Applied Mathematics and Computation} 219, pp. 63--77, 2012.

\bibitem{chaitin}
G.J. Chaitin, On the length of programs for computing finite binary
  sequences: statistical considerations.
\newblock {\em Journal of the ACM (JACM)} 16(1):145--159, 1969.

\bibitem{cover} T.M. Cover and J.A. Thomas, \emph{Elements of Information Theory}, 2nd Edition, Wiley-Blackwell, 2009.

\bibitem{ploscompbio} N. Gauvrit, H. Zenil, F. Soler-Toscano, J.-P. Delahaye, P. Brugger, Human Behavioral Complexity Peaks at Age 25, \textit{PLoS Comput Biol} 13(4): e1005408, 2017.

\bibitem{godel} K. G\"odel, \"Uber formal unentscheidbare S\"atze der Principia Mathematica und verwandter Systeme, I. \textit{Monatshefte f\"ur Mathematik und Physik} 38: pp. 173--98, 1931.

\bibitem{kamal} K. Dingle, C. Camargo, A.A. Louis, Input-output maps are strongly biased towards simple outputs, \textit{Nature Communications},  9(761), 2018.




\bibitem{kirchherr} W. Kirchherr, M. Li, and P. Vit\'anyi, `The miraculous universal distribution', \textit{Mathematical
Intelligencer}, Volume 19, Issue 4, (1997), pp 7--15.

\bibitem{kolmo} A.N.Kolmogorov, Three approaches to the quantitative definition of
  information.
\newblock {\em International Journal of Computer Mathematics} 2(1-4):157--168, 1968.

\bibitem{langton} C. Langton, Computation at the Edge of Chaos: Phase Transitions and Emergent Computation, In S. Forest (ed.) \textit{Emergent Computation}, The MIT Press, pp. 12--37, 1991.

\bibitem{levin}
L.A. Levin, Laws of information conservation (nongrowth) and aspects of the foundation of probability theory.
\newblock {\em Problemy Peredachi Informatsii} 10(3):30--35, 1974.

\bibitem{matiyasevich} Y.V. Matiyasevich,  Enumerable sets are Diophantine, \textit{Doklady Akademii Nauk SSSR} (in Russian). 191: 279--282. MR 0258744. English translation in \textit{Soviet Mathematics} 11 (2), pp. 354--357, 1970.

\bibitem{morzy} M. Morzy, T. Kajdanowicz, and P. Kazienko, On Measuring the Complexity of Networks: Kolmogorov Complexity versus Entropy, \emph{Complexity} vol. 2017.

\bibitem{post} E.L. Post, Finite Combinatory Processes -- Formulation 1, \textit{Journal of Symbolic Logic}, 1: pp. 103--105, 1936.

\bibitem{rado} T. Rad\'o, On non-computable functions, \textit{Bell System Technical Journal}, 41 (3): 877--884, 1962.

\bibitem{riedelzenil} J. Riedel, H. Zenil, Cross-boundary Behavioural Reprogrammability Reveals Evidence of Pervasive Universality, \textit{International Journal of Unconventional Computing}, vol 13:14-15 pp. 309--357, 2018.

\bibitem{riedelzenil2} J. Riedel and H. Zenil, Rule Primality, Minimal Generating Sets and Turing-Universality in the Causal Decomposition of Elementary Cellular Automata, \textit{Journal of Cellular Automata}, vol. 13, pp. 479--497, 2018.

\bibitem{d5} F. Soler-Toscano, H. Zenil, J.-P. Delahaye and N. Gauvrit, Calculating Kolmogorov Complexity from the Frequency Output Distributions of Small Turing Machines, \emph{PLoS ONE} 9(5): e96223, 2014.

\bibitem{solomonoff}
R.J. Solomonoff, A formal theory of inductive inference. parts i and ii.
\newblock {\em Information and control} 7(1):1--22 and 224--254, 1964.


\bibitem{wolfram} S. Wolfram, \emph{A New Kind of Science}, Wolfram Media, Champaign, IL, 2002.

\bibitem{wpp} S. Wolfram, \emph{A Class of Models with the Potential to Represent
Fundamental Physics} Volume 29, Issue 2, 2020.

\bibitem{zenilessay} H. Zenil. `The World is Either Algorithmic or Mostly Random', Third Prise Winning Essay -- Foundational Questions Institute (FQXi) Contest ``Is Reality Digital or Analog?''
2011.

\bibitem{liliana} H. Zenil, L. Badillo, S. Hern\'andez-Orozco and F. Hern\'andez-Quiroz, Coding-theorem Like Behaviour and Emergence of the Universal Distribution from Resource-bounded Algorithmic Probability, \textit{International Journal of Parallel Emergent and Distributed Systems}, 2018.

\bibitem{crc} H. Zenil, N.A. Kiani and J. Tegn\'er, Algorithmic Information Dynamics of Emergent, Persistent, and Colliding Particles in the Game of Life. In A. Adamatzky (ed), From Parallel to Emergent Computing (book)
Taylor \& Francis / CRC Press, pp.367--383, 2019.

\bibitem{kolmo2d} H. Zenil, F. Soler-Toscano, J.-P. Delahaye and N. Gauvrit, \emph{Two-Dimensional Kolmogorov Complexity and Validation of the Coding Theorem Method by Compressibility}, 2013.

\bibitem{bdmpaper} H. Zenil, F. Soler-Toscano, N.~A. Kiani, S. Hern\'andez-Orozco, and A. Rueda-Toicen.
\newblock A decomposition method for global evaluation of {S}hannon entropy and local estimations of algorithmic complexity. \newblock {\em 	arXiv:1609.00110 [cs.IT]}, 2016.

\bibitem{zenilreview2} H. Zenil, Towards Demystifying Shannon Entropy, Lossless Compression, and Approaches to Statistical Machine Learning, \textit{Proceedings of the International Society for Information Studies 2019 summit}, University of California, Berkeley, 2020 

\bibitem{zkpaper} H. Zenil, N.A. Kiani and J. Tegn\'er, Low Algorithmic Complexity entropy-deceiving Graphs, \textit{Physical Review E} 96, 012308, 2017.

\bibitem{randomness} H. Zenil, \textit{Randomness Through Computation: Some Answers, More Questions}, World Scientific Pub Co, Singapore, 2011.

\bibitem{computableuniverse} H. Zenil, \textit{A Computable Universe: Understanding and Exploring Nature as Computation}, World Scientific Pub Co, Singapore, 2012.

\bibitem{naturalcomputing} H. Zenil, Turing patterns with Turing machines: emergence and low-level structure formation, \textit{Natural Computing}, vol 12(4): 499--515, 2013.

\bibitem{blogpost} H. Zenil, Blog of \url{http://www.mathrix.org/liquid/archives/the-history-of-the-chaitin-leibniz-medallion} Accesed on February 26, 2019.


\bibitem{nmi} H. Zenil, N.A. Kiani, A. Zea, J. TegnÃ©r, Causal Deconvolution by Algorithmic Generative Models, \textit{Nature Machine Intelligence}, vol 1, pages 58--66, 2019.



\bibitem{frontiers} S. Hernández-Orozco, H. Zenil, J. Riedel, A. Uccello , N.A. Kiani, and J. Tegnér Algorithmic Probability-guided Machine Learning On Non-differentiable Spaces, \textit{Frontiers in Artificial Intelligence}, 25, 2021.

\bibitem{zenilruntimes} H. Zenil, From Computer Runtimes to the Length of Proofs: With an Algorithmic Probabilistic Application to Waiting Times in Automatic Theorem Proving In M.J. Dinneen, B. Khousainov, and A. Nies (Eds.), Computation, Physics and Beyond International Workshop on Theoretical Computer Science, WTCS 2012, LNCS 7160, pp. 223-240, Springer, 2012. 

\bibitem{zenilreview} H. Zenil, A Review of Methods for Estimating Algorithmic Complexity: Options, Challenges, and New Directions, \textit{Entropy}, 22, 612, 2020.

\bibitem{nar} H. Zenil, P. Minary, Training-free Measures Based on Algorithmic Probability Identify High Nucleosome Occupancy in DNA Sequences, \textit{Nucleic Acids Research}, gkz750, 2019.


\bibitem{thermo} H Zenil, C Gershenson, JAR Marshall, DA Rosenblueth, Life as thermodynamic evidence of algorithmic structure in natural environments, Entropy 14 (11), 2173-2191, 2012.

\bibitem{gaucherel} H Zenil, JP Delahaye, C Gaucherel, Image information content characterization and classification by physical complexity, \textit{Complexity}, 26-42,2010.

\bibitem{finance} H Zenil, JP Delahaye, An algorithmic information theoretic approach to the behaviour of financial markets, \textit{Journal of Economic Surveys} 25 (3), 431-463, 2011.

\bibitem{smalldata} H Zenil, Algorithmic data analytics, small data matters and correlation versus causation, \textit{Berechenbarkeit der Welt?}, 453-475, 2017

\bibitem{empirical} H Zenil, F Soler-Toscano, JJ Joosten, Empirical encounters with computational irreducibility and unpredictability, \textit{Minds and Machines} 22 (3), 149-165, 2012.

\bibitem{maininfo} H Zenil, NA Kiani, F Marabita, Y Deng, S Elias, A Schmidt, G Ball, J. Tegn\'er, An algorithmic information calculus for causal discovery and reprogramming systems, \textit{iScience} 19, 1160-1172, 2019.

\bibitem{dynamic} H Zenil, On the Dynamic Qualitative Behaviour of Universal Computation, \textit{Complex Systems} 20 (3), 265-278, 2012.

\bibitem{mutations} S Hern\'andez-Orozco, NA Kiani, H Zenil, Algorithmically probable mutations reproduce aspects of evolution, such as convergence rate, genetic memory and modularity, \textit{Royal Society open science}, 5 (8), 180399.

\bibitem{comprehension} H. Zenil, Compression is Comprehension, and the Unreasonable Effectiveness of Digital Computation in the Natural World. In S. Wuppuluri , F. Doria (eds.) \textit{Unravelling Complexity} (Gregory Chaitin's 70 festschrift), World Scientific Publishing, 2019.

\bibitem{scholarpedia} H Zenil, N.A. Kiani, F.S. Abrah{\~a}o, J. Tegn\'er, Algorithmic Information Dynamics, \textit{Scholarpedia}, 15(7):53143, 2020.


\end{thebibliography}

\end{document}